\documentclass[showpacs,preprintnumbers,aps]{revtex4}
\begin{document}
\preprint{USM-TH-202}
\title{Remarks on Axion-like models}
\author{Patricio Gaete}
\email{patricio.gaete@usm.cl}
\author{Iv\'an  Schmidt}
\email{ivan.schmidt@usm.cl} \affiliation{Center of Subatomic
Studies}
\affiliation{Department of Physics, Universidad T\'ecnica
Federico Santa Mar\'{\i}a, Valpara\'{\i}so, Chile}
\date{\today}

\begin{abstract}
For a recently proposed alternative to the traditional axion model,
we study its long distance behavior, in particular the confinement
versus screening issue, and show that a compactified version of this
theory can be further mapped into the massive Schwinger model. Our
calculation is based on the gauge-invariant but path-dependent
variables formalism. This result agrees qualitatively with the usual
axion model.
\end{abstract}
\maketitle

\section{Introduction}

Axion-like models \cite{Masso,Antoniadis,Heyl,Raffelt}, or simply
axions models, have become the focus of intense research activity
after recent results of the PVLAS collaboration \cite{Zavattini}.
This collaboration has reported measurements of the rotation of the
polarization of photons passing through a vacuum cavity in an
external magnetic field. As is well known, this effect can be
qualitatively understood by the existence of light pseudoscalars
bosons $\phi$ (the ''axion''), with a coupling to two photons. In
particular, the Lagrangian density of the photon-pseudoscalar system
is given by
\begin{equation}
{\cal L} =  - \frac{1}{4}F_{\mu \nu } F^{\mu \nu }  -
\frac{g}{8}\phi \varepsilon ^{\mu \nu \rho \sigma } F_{\mu \nu }
F_{\rho \sigma } + \frac{1}{2}\partial _\mu  \phi \partial ^\mu \phi
- \frac{{m_A^2 }}{2}\phi ^2,  \label{int5}
\end{equation}
where $m_A$ is the mass for the axion field  $\phi$. It is worth
recalling at this stage that this theory experiences mass generation
when the gauge field $F_{\mu \nu }$ takes a magnetic type
expectation value \cite{Spallucci}. If $F_{\mu \nu }$ takes an
electric type expectation value, tachyonic mass generation takes
place \cite{Spallucci}. Moreover, this theory leads to confining
potentials in the presence of nontrivial constant expectation values
for the gauge field strength $F_{\mu \nu}$ \cite{GaeteGuen}. In
particular, in the case of a constant electric field strength
expectation value the static potential remains Coulombic, while in
the case of a constant magnetic field strength expectation value the
potential energy is the sum of a Yukawa and a linear potential,
leading to the confinement of static charges. Notice that the
magnetic character of the field strength expectation value needed to
obtain confinement is in agreement  with the current chromo-magnetic
picture of the $QCD$ vacuum \cite{Savvidy}. Another feature of this
model is that it restores the rotational symmetry (in the
potential), despite of the fact that the external fields break this
symmetry. More interestingly, similar results have been obtained in
the context of the dual Ginzburg-Landau theory \cite{Suganuma}, as
well as for a theory of antisymmetric tensor fields that results
from the condensation of topological defects as a consequence of the
Julia-Toulouse mechanism \cite{GaeteW}. Accordingly, the above
interrelations are interesting from the point of view of providing
unifications among diverse models as well as exploiting the
equivalence in explicit calculations. We also point out that
confinement as a consequence of the interaction between a
non-Abelian constant chromo-magnetic background and the axion field
has been recently investigated in \cite{GaeteSpallu}.

On the other hand, we further observe that recently a novel way to
describe axions has been proposed \cite{Antoniadis}. The motivation
for this is mainly to reconcile the results of the PVLAS experiment
with both astrophysical bounds and the results of the CAST
collaboration. The crucial ingredient of this development is the
existence of a new light vector field ( rather than an axion field),
which interacts with the photon via Chern-Simons-like terms. In such
a case the Lagrangian density reads
\begin{equation}
{\cal L} =  - \frac{1}{4}F_{\mu \nu }^2 (A) - \frac{1}{4}F_{\mu \nu
}^2 (B)
 + \frac{{m_\gamma ^2 }}{2}A_\mu ^2  + \frac{{m_B^2 }}{2}B_\mu ^2  -
 \frac{{\kappa }}{4}\varepsilon ^{\mu \nu \lambda \rho } A_\mu  B_\nu
 F_{\lambda \rho } (A), \label{int10}
\end{equation}
where $m_\gamma$ is the mass of the photon, and $m_{B}$ represents
the mass for the gauge boson $B$. In particular, this alternative
theory exhibits an effective mass for the component of the photon
along the direction of the external magnetic field, exactly as it
happens with the theory (\ref{int5}).

Given its possible relevance in order to explain the discrepancy
coming from the CAST collaboration, in this work we wish to further
elaborate on the physical content of the model (\ref{int10}), in
particular its long distance structure. To this end we will study
the confinement versus screening issue and show that this theory can
be further mapped into the massive Schwinger model
\cite{Schwinger,GaeteS}. Our calculation is accomplished by making
use of the gauge-invariant but path-dependent variables formalism
along the lines of Ref. \cite{GaeteGuen}. This approach provides a
physically-based alternative to the usual Wilson loop approach,
where the usual qualitative picture of confinement in terms of an
electric flux tube linking quarks emerges naturally. As we shall
see, our analysis reveals that although both theories (\ref{int5})
and (\ref{int10}) lead to an effective mass for the photon, the
physical content is quite different. In other words, the confining
nature of the potential is lost. On the other hand, if the same
calculation is performed in the presence of two compact spacelike
dimensions, we again find a confining potential. In this way we
establish a new and peculiar connection with the Schwinger model, in
the hope that this will be helpful to understand better axion-like
models.

\section{Interaction energy}

We now examine the interaction energy between static point-like
sources for the model (\ref{int10}). This can be done by computing
the expectation value of the energy operator $H$ in the physical
state $|\Phi\rangle$ describing the sources, which we will denote by
$ {\langle H\rangle}_\Phi$.

Before proceeding with the determination of the interaction energy,
it is worthwhile noticing that the coupling for both theories
(\ref{int5}) and (\ref{int10}) is identical. In fact, as was first
mentioned in Ref. \cite{Antoniadis}, substituting $B_\mu$ by
$\partial_\mu \phi$ in (\ref{int10}), the theory under consideration
assumes the form
\begin{equation}
{\cal L} =  - \frac{1}{4}F_{\mu \nu }^2  + \frac{{m_\gamma ^2 }}{2}
+ \frac{1}{2}\partial _\mu  \phi \partial ^\mu  \phi  - \frac{\kappa
}{{4m_B }}\varepsilon ^{\mu \nu \lambda \rho } F_{\mu \nu }
F_{\lambda \rho } \phi. \label{int15}
\end{equation}
Thus we shall begin considering the following generating functional:
\begin{equation}
{\cal Z} = \int {{\cal D}\phi {\cal D}A\exp \left\{ {  i\int {d^4
x{\cal L}} } \right\}}, \label{int20}
\end{equation}
where the Lagrangian density is given by (\ref{int15}). We restrict
ourselves to static scalar fields, a consequence of this is that
$\triangle\phi=-\nabla^2\phi$, with
$\triangle\equiv\partial_\mu\partial^\mu$. It also implies that,
after performing the integration over $\phi$ in ${\cal Z}$, the
effective Lagrangian density reads
\begin{equation}
{\cal L} =  - \frac{1}{4}F_{\mu \nu }^2  + \frac{{m_\gamma ^2
}}{2}A_\mu ^2 - \frac{{\kappa ^2 }}{{32m_B^2 }}\varepsilon ^{\mu \nu
\lambda \rho } F_{\mu \nu } F_{\lambda \rho } \frac{1}{{\nabla ^2
}}\varepsilon ^{\alpha \beta \gamma \delta } F_{\alpha \beta }
F_{\gamma \delta } -A_{\mu}J^{\mu}, \label{int25}
\end{equation}
where $J^{\mu}$ is the external current of the test charges.
Furthermore, as was explained in \cite{GaeteGuen}, this expression
can be rewritten as
\begin{equation}
{\cal L} =  - \frac{1}{4}f_{\mu \nu }^2  + \frac{{m_\gamma ^2
}}{2}A_\mu ^2 - \frac{{\kappa ^2 }}{{8m_B^2 }}\varepsilon ^{\mu \nu
\alpha \beta } \left\langle {F_{\mu \nu } } \right\rangle
\varepsilon ^{\lambda \rho \gamma \delta } \left\langle {F_{\lambda
\rho } } \right\rangle f_{\alpha \beta } \frac{1}{{\nabla ^2
}}f_{\gamma \delta } -A_{\mu}J^{\mu}, \label{int30}
\end{equation}
where $\left\langle {F_{\mu \nu } } \right\rangle$ represents the
constant classical background. Here $f_{\mu \nu } =\partial _\mu
A_\nu -\partial _\nu A_\mu$ describes fluctuations around the
background. The above Lagrangian arose after using $ \varepsilon
^{\mu \nu \alpha \beta } \left\langle {F_{\mu \nu } } \right\rangle
\left\langle {F_{\alpha \beta } } \right\rangle=0$ (which holds for
a pure electric or a pure magnetic background). By introducing the
notation $\varepsilon ^{\mu \nu \alpha \beta }
 \left\langle{F_{\mu \nu } } \right\rangle  \equiv v^{\alpha \beta }
 $ and $\varepsilon ^{\rho \sigma \gamma \delta } \left\langle {F_{\rho
\sigma } } \right\rangle  \equiv v^{\gamma \delta }$, expression
(\ref{int30}) then becomes
\begin{equation}
{\cal L} =  - \frac{1}{4}f_{\mu \nu }^2  + \frac{{m_\gamma ^2
}}{2}A_\mu ^2 - \frac{{\kappa ^2 }}{{8m_B^2 }}v^{\alpha \beta }
f_{\alpha \beta } \frac{1}{{\nabla ^2 }}v^{\gamma \delta } f_{\gamma
\delta} - A_{\mu}J^{\mu}, \label{int35}
\end{equation}
where the tensor $v^{\alpha \beta }$ is not arbitrary, but satisfies
$\varepsilon ^{\mu \nu \alpha \beta } v_{\mu \nu } v_{\alpha \beta
}=0$.

\subsection{Magnetic case}

We now proceed to calculate the interaction energy in the $v^{0i}
\ne 0$ and $v^{ij}=0$ case (referred to as the magnetic one in what
follows). Using this in (\ref{int35}) we then obtain
\begin{equation}
{\cal L} =  - \frac{1}{4}f_{\mu \nu } f^{\mu \nu }  +
\frac{{m_\gamma ^2 }}{2}A_\mu ^2  - \frac{{\kappa ^2 }}{{8m_B^2
}}v^{0i} f_{0i} \frac{1}{{\nabla ^2 }}v^{0k} f_{0k}  - A_0 J^0,
\label{int40}
\end{equation}
with $(\mu ,\nu  = 0,1,2,3)$ and $(i,k= 1,2,3)$.

Now, we move on to compute the canonical Hamiltonian. For this end
we perform a Hamiltonian constraint analysis. The canonically
conjugate momenta are $\Pi^0=0$ and $\Pi _i  = D_{ij} E_j$, where
$E_i \equiv f _{i0}$ and $D_{ij}  \equiv \left( {\delta _{ij} -
\frac{{\kappa^2 }}{4m^2_B}v_{i0} \frac{1}{{ \nabla ^2}}v_{j0} }
\right)$. Since $D$ is a nonsingular matrix $(\det D = 1
-\frac{{\kappa^2 }}{4m_B^2}\frac{{{\bf v}^2 }}{{ \nabla ^2}} \ne 0)$
with ${\bf v}^2 \equiv v^{i0} v^{i0}$, there exists the inverse of
$D$. Solving for $E_{i}$ as a function of $\Pi_{i}$, we get
\begin{equation}
E_i  = \frac{1}{{\det D}}\left\{ {\delta _{ij} \det D +
\frac{{\kappa^2 }}{4m_B^2}v_i \frac{1}{{ \nabla ^2}}v_j }
\right\}\Pi _j. \label{int55}
\end{equation}

This leads us to the canonical Hamiltonian,
\begin{equation}
H_C  = \int {d^3 x} \left\{ { - A_0 \left( {\nabla\cdot {\bf \Pi} +
\frac{{m_\gamma ^2 }}{2}A^{0}  - J^0 } \right) + \frac{1}{2}{\bf
\Pi} ^2 + \frac{1}{2}{\bf B}^2  + \frac{{m_\gamma ^2 }}{2}{\bf A}^2
+ \frac{{\kappa ^2 }}{{8m_B^2 }}\frac{{\left( {{\bf v}\cdot{\bf \Pi}
} \right)^2 }}{{\left( {\nabla ^2  - \frac{{\kappa ^2 }}{{4m_B^2
}}{\bf v}^2 } \right)}}} \right\}, \label{int60}
\end{equation}
where ${\bf B}$ is the magnetic field. Requiring the primary
constraint
 $ \Pi_0=0$ to be preserved in time yields the following secondary constraint
\begin{equation}
\Gamma \left( x \right) \equiv \partial _i \Pi ^i  + m_\gamma ^2 A^0
- J^0=0. \label{int65}
\end{equation}
The above result reveals that there are two constraints, which are
second class. It explicitly reflects the breaking of the gauge
invariance of the theory under consideration. As a consequence,
special care has to be exercised since it is the gauge invariance
property that generally establishes unitarity and renormalizability
in most quantum field theoretical models. In order to convert the
second class system into first class we will adopt the procedure
described in Refs.\cite{Clovis,Rabin}. An important feature of this
development is that the new system still has the basic features of
the original one and has reestablished the gauge symmetry. As was
explained in Refs.\cite{Clovis,Rabin}, we enlarge the original phase
space by introducing a canonical pair of fields $\theta$ and $\Pi
_\theta$. Accordingly, a new set of constraints can be defined in
this extended space:
\begin{equation}
\Lambda _1  \equiv \Pi _0  + m_\gamma ^2  \theta  = 0, \label{int70}
\end{equation}
and
\begin{equation}
\Lambda _2  \equiv \Gamma  + \Pi _\theta   = 0. \label{int75}
\end{equation}
It can be easily checked that the new constraints are first class
and in this way restore the gauge symmetry of the theory under
consideration. It is worthwhile remarking at this point that the
$\theta$ fields only enlarge the unphysical sector of the total
Hilbert space, not affecting the structure of the physical subspace.
Therefore the new effective Lagrangian, after integrating out the
$\theta$, fields reads
\begin{equation}
{\cal L} =  - \frac{1}{4}f_{\mu \nu } \left( {1 + \frac{{m_\gamma ^2
}} {\Delta }} \right)f^{\mu \nu }  - \frac{{\kappa ^2 }}{{8m_B^2
}}v^{0i} f_{0i} \frac{1}{{\nabla ^2 }}v^{0k} f_{0k}  - A_0 J^0.
\label{int80}
\end{equation}

With this in hand, the canonical momenta are $ \Pi ^\mu   =  -
\left( {1 + \frac{{m_\gamma ^2 }}{\Delta }} \right)f^{0\mu }  -
\frac{{\kappa ^2 }} {{4m_B^2 }}v^{0\mu } \frac{1}{{\nabla ^2
}}v^{0i} f_{0i} $, and one immediately identifies the usual primary
constraint $\Pi^0=0$, and $\Pi _i  =  - \left( {1 + \frac{{m_\gamma
^2 }}{\Delta }} \right)f_{i0}  - \frac{{\kappa ^2 }}{{4m_B^2
}}v_{i0} \frac{1}{{\nabla ^2 }}v_{j0} f_{j0}$. The canonical
Hamiltonian can be worked out as usual and is given by the
expression
\begin{equation}
\begin{array}{r}
 H_C  = \int {d^3 x} \left\{ { - A_0 \left( {\partial _i \Pi ^i  - J^0 }
\right) + \frac{1}{2}\Pi ^i \left( {1 + \frac{{m_\gamma ^2 }}{\Delta
}} \right)^{ - 1} \Pi ^{i}  + \frac{1}{2}B^i \left( {1 +
\frac{{m_\gamma ^2 }}
{\Delta }} \right)B^{i} } \right\} +  \\
  + \int {d^3 x} \frac{{\kappa ^2 }}{{8m_B^2 }}\left( {{\bf v} \cdot \Pi }
\right)\frac{{\Delta ^2 }}{{\nabla ^2 \left( {\Delta  + m_\gamma ^2
} \right) \left[ {\left( {\Delta  + m_\gamma ^2 } \right) -
\frac{{\kappa ^2 {\bf v}^2 }}
{{4m_B^2 }}\frac{\Delta }{{\nabla ^2 }}} \right]}}\left( {{\bf v} \cdot \Pi } \right). \\
 \end{array} \label{int85}
\end{equation}

Temporal conservation of the primary constraint $\Pi_0$ leads to the
secondary constraint $\Gamma _1 \left( x \right) \equiv \partial _i
\Pi ^i - J^0 = 0$. The preservation in time of $\Gamma_1$ does not
give rise to any further constraints.  The extended Hamiltonian that
generates translations in time then reads $H = H_C + \int {d^3 }
x\left( {c_0 \left( x \right)\Pi _0 \left( x \right) + c_1 \left( x
\right)\Gamma _1 \left( x \right)} \right)$, where $c_0 \left( x
\right)$ and $c_1 \left( x \right)$ are the Lagrange multiplier
fields. Moreover, it is straightforward to see that $\dot{A}_0
\left( x \right)= \left[ {A_0 \left( x \right),H} \right] = c_0
\left( x \right)$, which is an arbitrary function. Since $ \Pi^0 =
0$ always, neither $ A^0 $ nor $ \Pi^0 $ are of interest in
describing the system and may be discarded from the theory. Thus the
Hamiltonian takes the form
\begin{equation}
\begin{array}{r}
 H_C  = \int {d^3 x} \left\{ {\frac{1}{2}\Pi ^i \left( {1 + \frac{{m_\gamma ^2 }}
 {\Delta }} \right)^{ - 1} \Pi ^{i}  + \frac{1}{2}B^i \left( {1 + \frac{{m_\gamma ^2 }}
 {\Delta }} \right)B^{i}  + c\left( x \right)\left( {\partial _i \Pi ^i  - J^0 }
 \right)} \right\} +  \\
  + \int {d^3 x} \frac{{\kappa ^2 }}{{8m_B^2 }}\left( {{\bf v} \cdot \Pi }
\right)\frac{{\Delta ^2 }}{{\nabla ^2 \left( {\Delta  + m_\gamma ^2
} \right) \left[ {\left( {\Delta  + m_\gamma ^2 } \right) -
\frac{{\kappa ^2 {\bf v}^2 }}
{{4m_B^2 }}\frac{\Delta }{{\nabla ^2 }}} \right]}}\left( {{\bf v} \cdot \Pi } \right), \\
\end{array}\label{int90}
\end{equation}
where $c(x) = c_1 (x) - A_0 (x)$.

According to the usual procedure we introduce a supplementary
condition on the vector potential such that the full set of
constraints becomes second class. A convenient choice is found to be
\cite{Pato}
\begin{equation}
\Gamma _2 \left( x \right) \equiv \int\limits_{C_{\xi x} } {dz^\nu }
A_\nu \left( z \right) \equiv \int\limits_0^1 {d\lambda x^i } A_i
\left( {\lambda x} \right) = 0, \label{int95}
\end{equation}
where  $\lambda$ $(0\leq \lambda\leq1)$ is the parameter describing
the spacelike straight path $ x^i = \xi ^i  + \lambda \left( {x -
\xi } \right)^i $, and $ \xi $ is a fixed point (reference point).
There is no essential loss of generality if we restrict our
considerations to $ \xi ^i=0 $. This allows us to write the only
nonvanishing equal-time Dirac bracket
\begin{equation}
\left\{ {A_i \left( x \right),\Pi ^j \left( y \right)} \right\}^ *
=\delta{ _i^j} \delta ^{\left( 3 \right)} \left( {x - y} \right) -
\partial _i^x \int\limits_0^1 {d\lambda x^j } \delta ^{\left( 3
\right)} \left( {\lambda x - y} \right). \label{int100}
\end{equation}

Finally we are ready to tackle the question of the interaction
energy between pointlike sources, where a fermion is localized at
${\bf y}\prime$ and an antifermion at ${\bf y}$. Since the fermions
are taken to be infinitely massive (static) we can substitute
$\Delta$ by $-\nabla^{2}$ in Eq. (\ref{int90}). In such a case we
write $\left\langle H \right\rangle _\Phi$ as
\begin{equation}
\left\langle H \right\rangle _\Phi   = \left\langle \Phi
\right|\int
 {d^3 x} \left\{ {\frac{1}{2}\Pi ^i \frac{{\nabla ^2 }}{{\nabla ^2  - M^2 }}
 \Pi ^i  + \frac{1}{2}B^i \left( {1 - \frac{{m_\gamma ^2 }}{{\nabla ^2 }}}
 \right)B^i } \right\}\left| \Phi  \right\rangle, \label{int105}
\end{equation}
with $ M^2  \equiv m_\gamma ^2  + \frac{{\kappa ^2 }}{{4m_B^2 }}{\bf
v}^2 = m_\gamma ^2  + \frac{{{\bf B}^2 }}{{v_B^2 }}$, where we have
employed ${\bf v}^{2} = 4{\bf B}^{2}$ and $\kappa  = \frac{{m_B
}}{{v_B }}$.

Next, as was first established by Dirac \cite{Dirac2}, the physical
state can be written as
\begin{equation}
\left| \Phi  \right\rangle  \equiv \left| {\overline \Psi  \left(
\bf y \right)\Psi \left( {\bf y}\prime \right)} \right\rangle  =
\overline \psi \left( \bf y \right)\exp \left( {ie\int\limits_{{\bf
y}\prime}^{\bf y} {dz^i } A_i \left( z \right)} \right)\psi
\left({\bf y}\prime \right)\left| 0 \right\rangle, \label{int110}
\end{equation}
where $\left| 0 \right\rangle$ is the physical vacuum state and the
line integral appearing in the above expression is along a spacelike
path starting at ${\bf y}\prime$ and ending at $\bf y$, on a fixed
time slice. It is worth noting here that the strings between
fermions have been introduced in order to have a gauge-invariant
function $\left| \Phi  \right\rangle $. In other terms, each of
these states represents a fermion-antifermion pair surrounded by a
cloud of gauge fields sufficient to maintain gauge invariance.

From the foregoing Hamiltonian discussion, we first observe that
\begin{equation}
\Pi _i \left( x \right)\left| {\overline \Psi  \left( \bf y
\right)\Psi \left( {{\bf y}^ \prime  } \right)} \right\rangle  =
\overline \Psi  \left( \bf y \right)\Psi \left( {{\bf y}^ \prime }
\right)\Pi _i \left( x \right)\left| 0 \right\rangle  + e\int_ {\bf
y}^{{\bf y}^ \prime  } {dz_i \delta ^{\left( 3 \right)} \left( {\bf
z - \bf x} \right)} \left| \Phi \right\rangle. \label{int115}
\end{equation}
Substituting this in (\ref{int105}), we get
\begin{equation}
\left\langle H \right\rangle _\Phi   = \left\langle H \right\rangle
_0  - \frac{{q^2 }}{{4\pi }}\frac{{e^{ - ML} }}{L}, \label{int120}
\end{equation}
where $\left\langle H \right\rangle _0  = \left\langle 0
\right|H\left| 0 \right\rangle$ and $|{\bf y}-{\bf
y}^{\prime}|\equiv L$. Since the potential is given by the term of
the energy which depends on the separation of the two fermions, from
the expression (\ref{int120}) we obtain
\begin{equation}
V= - \frac{{q^2 }}{{4\pi }}\frac{{e^{ - ML} }}{L}. \label{int125}
\end{equation}
As already stated, $M$ is the effective mass for the component of
the photon along the direction of the external magnetic field.
Accordingly, the above analysis reveals that, although both theories
(\ref{int5}) and (\ref{int10}) contain the same coupling, the
physical content is quite different. It is important to realize that
expression (\ref{int125}) is spherically symmetric, although the
external fields break the isotropy of the problem in a manifest way.
Another example where the rotational symmetry is restored was
studied in the case of non commutative QED \cite{G/S}.

\subsection{Electric case}

We shall now consider the case $v^{0i}=0$ and $v^{ij}\ne 0$
(referred to as the electric one in what follows). In such a case
the density Lagrangian becomes
\begin{equation}
{\cal L} =  - \frac{1}{4}f_{\mu \nu }^2  + \frac{{m_\gamma ^2 }}{2}
A_\mu ^2  - \frac{{\kappa ^2 }}{{8m_B^2 }}v^{ij} f_{ij} \frac{1}
{{\nabla ^2 }}v^{kl} f_{kl} , \label{int126}
\end{equation}
$(\mu ,\nu  = 0,1,2,3)$ and $(i,j,k,l = 1,2,3)$.

The above Lagrangian will be the starting point of the Dirac
constrained analysis. As before, the above Lagrangian leads to a
second class constraint. In view of this situation and following our
earlier procedure, the new effective Lagrangian takes the form
\begin{equation}
{\cal L} =  - \frac{1}{4}f_{\mu \nu } \left( {1 + \frac{{m_\gamma ^2
}} {\Delta }} \right)f^{\mu \nu }  - \frac{{\kappa ^2 }}{{8m_B^2
}}v^{ij} f_{ij} \frac{1}{{\nabla ^2 }}v^{kl} f_{kl} , \label{int127}
\end{equation}

The canonical momenta following from Eq.(\ref{int127}) are $\Pi ^\mu
=
  - \left( {1 + \frac{{m_\gamma ^2 }}{\Delta }} \right)f^{0 \mu } $, which
results in the usual primary constraint $\Pi^0=0$ and $\Pi ^i   =  -
\left( {1 + \frac{{m_\gamma ^2 }}{\Delta }} \right)f^{0 i}$.
Defining the electric and magnetic fields by $ E^i  = f^{i0}$ and
$B^i  = \frac{1}{2}\varepsilon ^{ijk} f_{jk}$, respectively, the
canonical Hamiltonian assumes the form
\begin{equation}
\begin{array}{r}
 H_C  = \int {d^3 x} \left\{ {\frac{1}{2}E_i \left( {1 + \frac{{m_\gamma ^2 }}
{\Delta }} \right)E^i  + \frac{1}{2}B_i \left( {1 + \frac{{m_\gamma
^2 }} {\Delta }} \right)B^i  + \frac{{\kappa ^2 }}{{8m_B^2
}}\varepsilon _{ijm}
\varepsilon _{k\ln } v^{ij} B^m \frac{1}{{\nabla ^2 }}v^{kl} B^n } \right\} -  \\
  - \int {d^3 x} A_{0}\left\{ {\partial _i \Pi ^i  - J^0 } \right\}. \\
 \end{array} \label{int128}
\end{equation}
Time conservation of the primary constraint leads to the secondary
constraint $\Gamma_1(x) \equiv \partial_i\Pi^i - J^0=0$, and the
time stability of the secondary constraint does not induce more
constraints, which are first class.  Following our earlier
procedure, we will compute the expectation value of the Hamiltonian
in the physical state $\left| \Phi \right\rangle$, that is,
\begin{equation}
\left\langle H \right\rangle _\Phi   = \left\langle \Phi
\right|\int {d^3 x} \left\{ {\frac{1}{2}E_i \left( {1 -
\frac{{m_\gamma ^2 }}{{\nabla ^2 }}}
 \right)E^i } \right\}\left| \Phi  \right\rangle.  \label{int129}
\end{equation}
Hence we see that the potential is given by
\begin{equation}
V =  - \frac{{q^2 }}{{4\pi }}\frac{{e^{ - m_\gamma  L} }}{L},
\label{int130}
\end{equation}
where $L\equiv|{\bf y}-{\bf {y^\prime}}|$. It must be observed that
in this case $M$ reduces to $m_\gamma$, and in the limit  $m_\gamma
\to 0$ the potential (\ref{int130}) reduces to the Coulomb one
\cite{GaeteGuen}.

\section{Final Remarks}

Our discussion, so far, has concentrated on the confinement versus
screening issue for the recently proposed axion model (\ref{int10}).
As we have seen, the coupling for this theory is identical to the
traditional axion model (\ref{int5}). We have also seen that this
alternative theory exhibits an effective mass for the component of
the photon along the direction of the external magnetic field, as it
happens with the theory (\ref{int5}). From this point of view, it is
meaningful to ask wether the confining nature of the potential can
be recovered in some approximations. We now address this question.

To that end, we will discuss the mapping of the theory (\ref{int35})
into the massive Schwinger model. As was explained in \cite{GGS}, we
illustrate this by making a dimensional compactification (\`a la
Kaluza-Klein) on Eq.(\ref{int35}). Then we see that the new theory
takes the form:
\begin{equation}
{\cal L}^{\left( {1 + 1} \right)}  =  - \frac{1}{4}f_{\mu \nu }
\sum\limits_n {\left( {1 + \frac{{\zeta ^2 }}{{\Delta _{\left( {1 +
1}
 \right)}  + a^2 }}} \right)} f^{\mu \nu }  - A_\mu  J^\mu , \label{int135}
\end{equation}
where $ \zeta ^2  = m_\gamma ^2  + \frac{{\kappa ^2 }}{{8m_B^2 }}
\left[ {g^{\alpha \mu } g^{\beta \nu } v_{\alpha \beta } g_{\gamma
\mu }
 g_{\delta \nu } v^{\gamma \delta } } \right]$ and $
a^2  \equiv {\raise0.7ex\hbox{${n^2 }$} \!\mathord{\left/
 {\vphantom {{n^2 } {R^2 }}}\right.\kern-\nulldelimiterspace}
\!\lower0.7ex\hbox{${R^2 }$}}$. We immediately recognize the above
to be the massive Schwinger model with mass $m^{2} \equiv a^{2}$.
Notwithstanding, in order to put our discussion into context it is
useful to summarize the relevant aspects of the analysis described
previously \cite{GGS}. We shall begin by recalling the interaction
energy for the massive Schwinger model taking a contribution of a
single mode in Eq.(\ref{int135}). We obtain \cite{GaeteS}:
\begin{equation}
V = \frac{{q^2 }}{{2\lambda }}\left( {1 + \frac{{a^2 }}{{\lambda ^2
}}} \right)\left( {1 - e^{ - \lambda L} } \right) + \frac{{q^2
}}{2}\left( {1 - \frac{{4 \zeta^2 }}{{\lambda ^2 }}} \right)L,
\label{int140}
\end{equation}
where $\lambda ^2  \equiv 4 \zeta^2  + a^2$ and $|{y}|\equiv L$.
Effectively, therefore, our initial theory (\ref{int35}) is mapped
into the massive Schwinger model, which displays both the screening
and the confining part of this interaction. Of course, if we
consider the zero mode case, i. e. $a=0$, the static potential above
shows that confinement disappears. As in \cite{GGS}, we will
concentrate on the second term of Eq. (\ref{int140}), which
represents confinement. The expression for the coefficient of the
linear potential between two static point sources is:
\begin{equation}
T = \frac{{q^2 }}{2}\sum\limits_{n_1 ,n_2 } {\frac{{\left(
{\frac{{n_1^2 }}{{R_1^2 }} + \frac{{n_2^2 }}{{R_2^2 }}} \right)}}{{{
\zeta^2 } + \frac{{n_1^2 }}{{R_1^2 }} + \frac{{n_2^2 }}{{R_2^2
}}}}}. \label{int145}
\end{equation}
Following our earlier procedure, in the limit $R_1,R_2 \to \infty$
we obtain
\begin{equation}
T = \pi q^2 R_1 R_2 \int_0^\Lambda  {d\rho \frac{{\rho ^3
}}{{{\zeta^2 } + \rho ^2 }}}, \label{int150}
\end{equation}
that is,
\begin{equation}
T = \frac{{\pi q^2 }}{2}R_1 R_2 \left[ {\Lambda ^2  -  \zeta^2 \ln
\left( {\frac{{ \zeta^2  + \Lambda ^2 }}{{ \zeta^2 }}} \right)}
\right], \label{int155}
\end{equation}
again if $R_1,R_2  \to \infty$, we obtain a  transcendental equation
for ${\raise0.7ex\hbox{$\Lambda $} \!\mathord{\left/
 {\vphantom {\Lambda  {C^2 }}}\right.\kern-\nulldelimiterspace}
\!\lower0.7ex\hbox{${ \zeta^2 }$}}$:
\begin{equation}
\frac{{\Lambda ^2 }}{{ \zeta^2 }} - \ln \left( {1 + \frac{{\Lambda
^2 }}{{\zeta^2 }}} \right) = 0. \label{int160}
\end{equation}
From (\ref{int150})  here we can deduce that as $\frac{{\Lambda ^2
}}{{\zeta^2 }} \sim \sqrt {\frac{{2\pi T}}{{q^2 R_1 R_2 }}} \to 0$,
which means that  $\zeta$ has to grow stronger than $\Lambda$ when
$\Lambda  \to \infty$, in order to obtain a finite coefficient of
the linear potential. It is clear from this discussion that our
phenomenological result (\ref{int140}) agrees qualitatively with the
magnetic case of the usual axion model \cite{GaeteGuen}, in the
limit of large L. Thus, only after the compactification, both
theories are equivalent in the low energy regime. In this way we
establish an intriguing analogy with the massive Schwinger model,
which simulates the features of the usual $(3+1)$-dimensional axion
model.

\section{ACKNOWLEDGMENTS}

P. G. was partially supported by Fondecyt (Chile) grant 1050546.

\end{document}